\documentclass[a4paper,11pt]{article}
\usepackage[T1]{fontenc} 
\usepackage[utf8]{inputenc}
\usepackage[english]{babel}
\usepackage{graphicx} 
\usepackage{authblk}
\usepackage{placeins}
\usepackage{subcaption}
\usepackage{mathrsfs}
\usepackage[dvipsnames]{xcolor}
\usepackage{enumerate} 
\usepackage{amsmath}
\usepackage{slashed}
\usepackage{color}
\usepackage[hidelinks]{hyperref}

\usepackage{amsfonts}
\usepackage{amssymb}
\usepackage{mathrsfs} 
\usepackage{cases} 
\usepackage{placeins} 
\usepackage{mathtools}

\newcommand{\be}{\begin{equation}}
\newcommand{\ee}{\end{equation}}
\newcommand{\bea}{\begin{eqnarray}}
\newcommand{\eea}{\end{eqnarray}}

\newcommand{\LG}{\mathcal{L}}

\title{\textcolor{RoyalPurple}{Large loop-coupling enhancement of a heavy pseudoscalar from a light dark sector}}
\date{\vspace{-5ex}}

\author[a]{Stefano Di Chiara}
\author[a]{Andi Hektor}
\author[a]{Kristjan Kannike}
\author[a,b]{\\Luca Marzola}
\author[a,b]{Martti Raidal}

\affil[a]{\normalsize\it National Institute of Chemical Physics and Biophysics; R\"{a}vala pst. 10\\ 10143 Tallinn, Estonia}
\affil[b]{\normalsize\it Institute of Physics, University of Tartu; Ravila 14c, 50411 Tartu, Estonia}

\begin{document}
\maketitle 

\begin{abstract}
The small background and the sensitivity to charged particles via a leading order loop coupling make the diphoton channel a privileged experimental test for new physics models. We propose a simple archetypal scenario to generate a sharp di-photon resonance as a result of threshold enhancements in the effective coupling between a heavy pseudoscalar particle and new vector-like leptons. We therefore study three different scenarios consistent with the current experimental limits and deviating from the Standard Model at the 2~$\sigma$ level. The model also introduces a natural dark matter candidate able to match the observed dark matter abundance and comfortably respect the current direct detection constraints.

\end{abstract}

\section{Introduction}
\label{intro}

Last year both the ATLAS and CMS experiments of the Large Hadron Collider (LHC) at CERN reported the presence of an excess in the diphoton channel, peaked at a centre-of-mass energy of about 750~GeV~\cite{diphotonATLAS, diphotonCMS}. The signal appeared with a statistical significance of about 2.6$\sigma$ in the data gathered by the CMS detector, while the ATLAS collaboration measured a 3.6$\sigma$ excess. The two indications, being compatible within the limits of the resolutions of the detectors, triggered an incredible number of works elaborating on the possible beyond-the-Standard-Model origin of the excess.\footnote{A comprehensive collection of the papers on the topic is presented in \cite{Staub:2016dxq}. }
In March 2016, both the ATLAS and CMS collaborations updated their analysis with the new data collected at 13 TeV. 
The observations based on an integrated luminosity of 3.3~fb${}^{-1}$ hinted once again at the particle interpretation of the signal, with an excess reaching a local significance of $3.4\sigma$ and $3.9\sigma$ in the data of CMS and ATLAS, respectively \cite{cmsmoriond,CMS:2016owr,atlasmoriond}. Disappointingly, though, the significance of the purported signal was seriously impaired in a later analysis of larger datasets by the ATLAS and CMS groups, respectively relying on 15.4 and 12.9~fb${}^{-1}$ of acquired data \cite{ATLAS:2016eeo,CMS:2016crm}. In spite of the fate of the $750$ GeV resonance, the episode still exemplifies the potential of the diphoton final state: such a channel is indeed an important tool for searches of heavy neutral spin-zero resonances owing to the smallness of the involved background and the sensitivity to new physics via a leading order one loop contribution. 
In this regard, we remark that even the 125~GeV Higgs boson was first signalled by a resonance in the same diphoton channel. Furthermore, the latest LHC data present deviations at the one to two $\sigma$ level associated to energies larger smaller or equal to about a TeV which might be confirmed in forthcoming analyses. 
We thus find compelling to study how potential signals of diphoton resonances are entwined with otherwise unobserved new physics states which, within particle models that are well defined up to very high energies, provide their effective coupling to the photons.

According to the Landau-Yang theorem, a resonance in the diphoton channel can originate only from a particle of either spin zero or two. In the remaining of the paper we assume that the speculated particle has spin zero and investigate a way to produce a signal large enough for the corresponding resonance to be possibly discoverable in a dataset of the order of hundreds of fb${}^{-1}$,\footnote{The integrated luminosity projected to be delivered by LHC at the end of Run II is about 200~fb${}^{-1}$.} but still consistent with the current experimental limits. As it was the case for the 750~GeV excess, we require the absence of indications in complementary channels such as the di-jet, the $t\bar t$, the di-boson and the di-lepton ones. Interestingly, this condition necessarily forces the introduction of additional charged and colored particles characterized by large multiplicities and/or large couplings to the new resonance. However, assuming a diphoton cross section of $O(\rm{fb})$, the running of these couplings seems to drive the model to a non-perturbative regime already at scales as low as a few TeVs \cite{Franceschini:2015kwy, Bertuzzo:2016fmv}. While this fact apparently favours frameworks supporting the compositeness of the hypothetical spin-zero particle, we aim to demonstrate that a perturbative scenario based on a fundamental spin-0 field is still attainable. 

In the following we study three cases of fundamental spin-0 resonance characterized by masses of 330~GeV, 720~GeV, and 1000~GeV, respectively, and show that a large excess in diphoton events can be comfortably reproduced within a simple extension of the Standard Model (SM) which retains perturbativity up to scales at least as large as $O(\rm 10^{10})$~TeV and up to the Planck scale in the best case of the three scenarios. In the framework we consider in this study, the high mass diphoton resonance originates from a new pseudoscalar particle that couples to a vector-like (VL) charged lepton, characterized by a mass not far from half the mass of the new scalar resonance, and to two VL neutrinos. The pseudoscalar production at the LHC is allowed by the coupling to a heavier VL top quark that mediates the gluon fusion process. Our goal is to demonstrate that, in this setup, the new particle content needs only modest Yukawa couplings to generate the large diphoton cross section initially observed at the LHC. 

Interestingly, in our analysis we also find that the lighter VL neutrino is a viable dark matter (DM) candidate and that, in the same parameter space that yields a sizeable resonant signal, this particle gives rise to a DM abundance in the ballpark of the measured one. The scenario also complies with the present direct detection constraints.

The paper is organized as follows: in Section~\ref{FM} we review the motivation that lead to the choice of our framework, which is detailed in Section~\ref{model}.
The relevant LHC phenomenology is discussed in Section~\ref{sec:The LHC di-photon excess} whereas in Section~\ref{sec:DM} we show that the proposed VL neutrino is a viable DM candidate. We gather our conclusions in Section~\ref{sec:Conclusions}.

\section{Threshold enhancement of the diphoton decay rate}
\label{FM}

As a prototypal case we study here a possible LHC diphoton signal with 720~GeV invariant mass. For the relevant centre-of-mass energies of 8 and 13 TeV, the largest contribution to the production cross section is given by the gluon-gluon fusion process. Hence, we introduce additional colored and charged particles to provide the effective coupling between the SM gauge bosons and the speculated 720~GeV particle via loops. The contributions of the new states to the diphoton decay width are inversely proportional to the square of the masses of the new particles. Since any new charged and colored vector bosons must be much heavier than 720~GeV because of the present experimental bounds, it follows that the contribution of new spin-1 particles to the effective coupling is necessarily suppressed. As for the remaining possibilities, the partial amplitude mediated by a heavy fermion loop is four times as large as that mediated by a heavy scalar, and therefore fermions are more suitable in generating a large diphoton cross section. 

In the present scenario we then consider VL fermions $f$ (whose mass terms, contrarily to those of chiral fermions, are gauge invariant) characterized by a charge $e_f$, in units of the positron charge, and $N_f$ colors. The contribution to the diphoton decay width of a scalar $H$ or of a pseudoscalar $A$ of these VL fermions is then quantified by \cite{Djouadi:2005gj,Gunion:1989we}
\bea\label{dip}
\Gamma_{S\rightarrow \gamma\gamma}&=& \frac{\alpha_e^2 m_{S}^3}{256 \pi^3 v_w^2}\left| \sum_f a_f  N_f e^2_f F^S_{f} \right|^2\,, \qquad S\in\{A,H\}
\eea
where
\be
a_f=\frac{y_f v_w}{\sqrt{2} m_f}\,,\qquad \tau_{f}=\frac{4 m_f^2}{m_{S}^2}\,, 
\ee
and
\be
F^H_{f}=-2 \tau_{f}\left[1+\left( 1-\tau_{f} \right) f(\tau_{f})\right]\,,
\qquad F^A_{f}=-2 \tau_{f} f(\tau_{f})\, .
\ee
The pseudoscalar partial amplitude is affected by a discontinuity, such that
\be
\lim_{m_f\to \left(m_A/2\right)^\pm} F^A_{f}=\frac{\pi^2}{2}\,,\ F^A_{f}|_{m_f = m_A/2}=\pi^2\,,
\ee
which originates from threshold effects. We avoid this discontinuity in our computation by setting $F^A_{f}=\pi^2/2$ at the threshold.

\begin{figure}[ht]
  \centering
    \includegraphics[width=.48\linewidth]{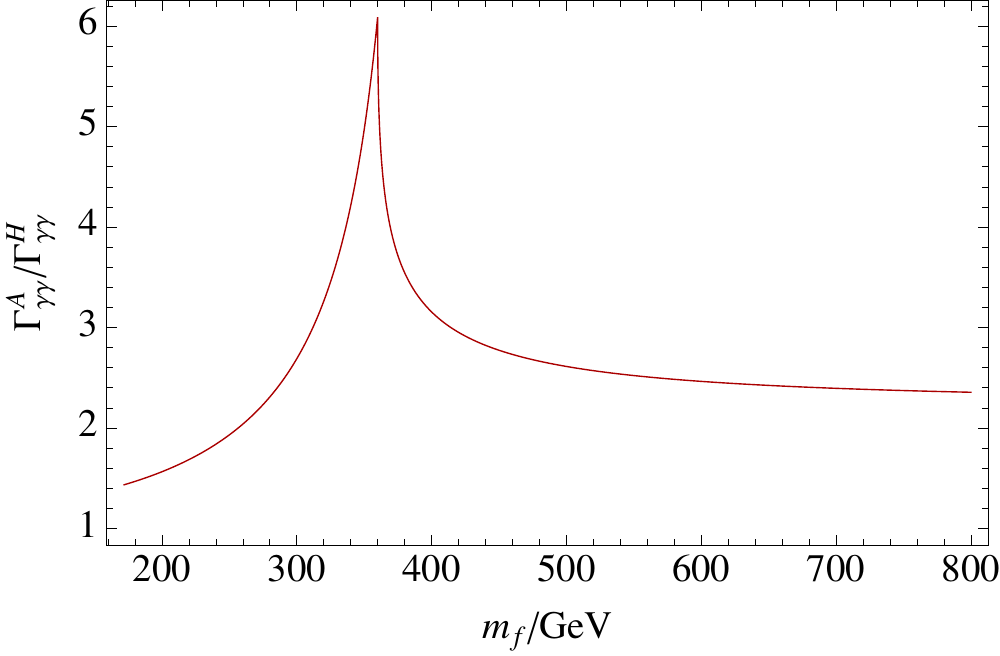}
    \includegraphics[width=.48\linewidth]{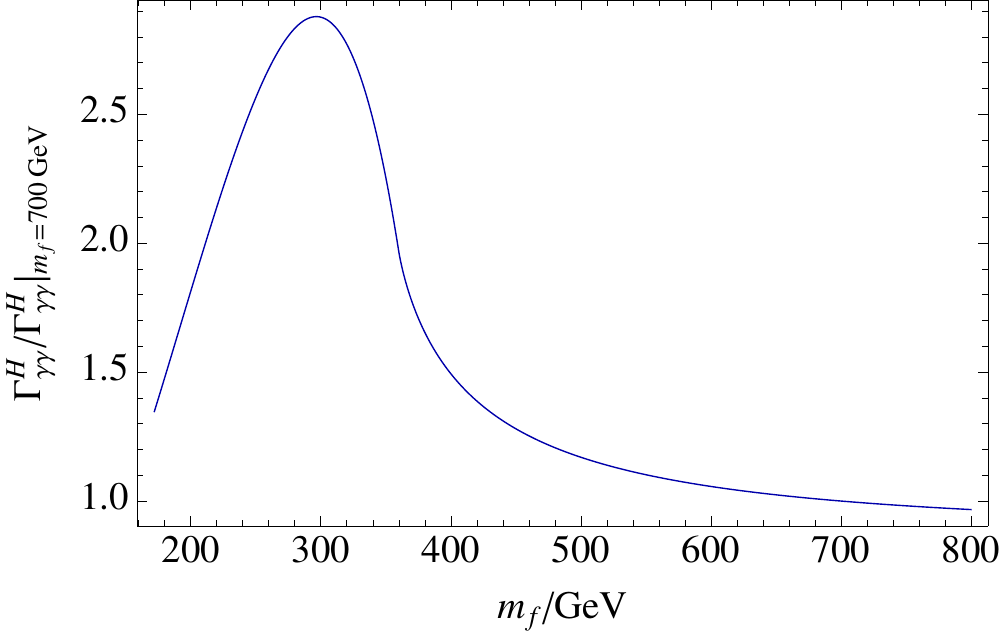}
  \caption{Left panel: Ratio of the diphoton decay rate of a pseudoscalar $A$ to that of a scalar $H$ mediated by a VL fermion loop. Right panel: the scalar decay rate to diphoton as a function of VL fermion mass $m_f$, normalized to the case of $m_f=700$~GeV.}
  \label{GAH}
\end{figure}

In the left panel of Fig.~\ref{GAH} we show the diphoton decay rates of a 720~GeV pseudoscalar particle $A$, via a VL fermion loop, normalized to the corresponding quantity for a scalar $H$. As one can see, the large enhancement of the loop coupling of $A$ to photons for $m_f=m_A/2$ allows the pseudoscalar decay width to match the scalar one through a Yukawa coupling between $A$ and the VL fermion that is about 2.5 times smaller than that of $H$. In the right-hand side panel of Fig.~\ref{GAH}, we furthermore show how the partial amplitude for the diphoton decay of $H$ depends on $m_f$. The normalization here is given by the corresponding quantity computed for $m_f=700$~GeV, which approximately matches the current lower bound on the mass of heavy VL quarks \cite{CMS:1900uua}. 
We then conclude that, to achieve a given decay rate, the Yukawa coupling of a 720~GeV pseudoscalar to a 365~GeV charged VL lepton must be about 3.5 times smaller than that required by a scalar which couples to photons via a 700~GeV VL quark loop.

Motivated by the observations above, we identify the speculated 720~GeV particle with a CP-odd scalar $A$ that couples only to VL fermions. The relevant diphoton cross section, in the narrow width approximation, is then given by
\be\label{sppSgg}
\sigma\left(p p\rightarrow A\rightarrow\gamma\gamma \right)=\sigma^{\rm SM}_{pp\rightarrow H}\frac{\Gamma_{gg}}{\Gamma^{\rm SM}_{gg}}\frac{\Gamma_{\gamma\gamma}}{ \Gamma_{\text{tot}}}\,,
\ee
where the SM labeled quantities, respectively the LHC cross section at 13~TeV for the production via digluon fusion of a SM Higgs boson of mass $m_A$ and the decay rate of the same boson to digluon, are tabulated in \cite{Dittmaier:2011ti,Dittmaier:2012vm,Heinemeyer:2013tqa}. The decay rate of the pseudoscalar $A$ to digluons in Eq.~\eqref{GAH} is defined by \cite{Djouadi:2005gj,Gunion:1989we} 
\be\label{dig}
\Gamma_{A\rightarrow g g}= \frac{\alpha_s^2 m_{A}^3}{128 \pi^3 v_w^2}\left| \sum_f a_f F^A_{f} \right|^2\,,
\ee
with $f=T$, while $\Gamma_{\text{tot}}$ is the total decay rate of $A$. The cross section in Eq.~\eqref{sppSgg} is maximized for $\Gamma_{\text{tot}}\sim\Gamma_{gg}$, which in the present scenario holds as long as the Yukawa couplings of the VL quarks are comparable to those of the charged VL leptons. The lower limit for heavy copies of SM quarks decaying to a top or bottom quark and a $W^\pm$, or a SM Higgs, ranges from 705~GeV to 846~GeV \cite{CMS:1900uua, CMS:2014cca}. These can be relaxed to 690~GeV in case of decays to light quarks \cite{Aad:2015tba}. Charged heavy leptons, depending on their charge, must instead be heavier than at least 400~GeV \cite{Chatrchyan:2013oca}. This limit is relaxed to 104~GeV for charged particles of an SU$(2)_L$ doublet which decays to a nearly degenerate neutral component \cite{Abbiendi:2002vz}. The latter, being weakly interacting and stable, is a viable DM candidate.  Motivated by this intriguing example, in the next section we introduce a model that captures and generalises the features of the diphoton resonance we discussed above.

\section{The model}
\label{model}

In order to model potential signals of resonances appearing exclusively in the diphoton channel, we extend the SM particle content with a VL lepton EW doublet and singlet and a VL top quark, which by construction do not contribute to the anomaly diagrams of the $SU(3)_c \times SU(2)_L \times U(1)_Y$ gauge group. To avoid lepton-number violating processes we also impose a $\mathbb{Z}_2$ symmetry at the Lagrangian level. In particular, we assume that all the VL fermions are odd under the $\mathbb{Z}_2$, while the SM leptons and $3^{\rm rd}$ generation quarks, as well as the Higgs boson and the pseudoscalar $A$, are even. The first two generations quarks are also taken odd under the discrete symmetry so that the VL quark is allowed to decay into these states via a small Yukawa coupling \cite{Ge:2016xcq}.

\begin{table}[htp]
\caption{Scalar and VL fermion content of the model. Our convention for the electric charge is $Q=Y+I_3$. We denote with $H$ the usual SM Higgs doublet.}
\begin{center}	
\begin{tabular}{|ccccc|}
\hline
Field & $SU(3)_c $ & $SU(2)_L$ & $U(1)_Y $ & $\mathbb{Z}_2$\\
\hline
 $H$ & 1 & $\begin{pmatrix} \phi^+ \\ \left(v_w+h+i \phi^0\right)/\sqrt{2} \end{pmatrix}$ & 1/2 & $+$ \\
 $A$ &  1&  $A$  & 0 & $+$ \\
 $L$ & 1 &  $\begin{pmatrix} N \\ E \end{pmatrix}$ & $-1/2$ & $-$\\
 $N'$& 1 & $N'$ & $0$ & $-$ \\
 $T$ & 3& $T$ & $2/3$ & $-$\\
 \hline
\end{tabular}
\end{center}
\label{4thgen}
\end{table}%

Finally, we assume the Lagrangian to respect CP symmetry, which forbids linear and cubic terms in the $A$ potential:

\begin{align}\label{L}
	\LG \supset
	 &\left[
	y^L_N \bar{L}_L \tilde{H} N'_R +y^R_N \bar{L}_R \tilde{H} N'_L+
	\text{H.c.}\right]
	-
	i y_L A  \bar{L}\gamma^5 L-i y_N A \bar{N'} \gamma^5 N'-i y_T A \bar{T} \gamma^5 T
	\nonumber\\
	&+
	m_L \bar{L} L
	+
	m_{N} \bar{N'} N'
	+		
	m_{T} \bar{T} T
	-
	m_A^2 A^2
	-
	\lambda_A A^4
	-
	\lambda_{AH} A^2 |H|^2
	\,.
\end{align}
Here we take a positive portal coupling, $\lambda_{AH}>0$, that prevents $A$ from acquiring a CP-violating vacuum expectation value and omitted the small Yukawa couplings of the VL quark to the light quarks. 

We remark that VL fermions coupled to heavy spin-zero resonances appear in a class of string-inspired supersymmetric models \cite{Cvetic:2015vit,Dev:2015vjd,King:2016wep,Karozas:2016hcp}, of which our framework possibly is an effective low energy limit.

In Appendix~\ref{msmx} we present the masses and pseudoscalar couplings of the VL neutral leptons. The mass assignments for our VL fermions allow the charged component $E$ to decay into the lighter VL neutrino via the process $E\rightarrow N_1+W^{\pm*}\rightarrow N_1+ l^\pm\nu_l$. A small mass splitting between the charged and the neutral VL leptons fulfils the requirement in \cite{Abbiendi:2002vz} and respects the constraint from the $T$ parameter \cite{Peskin:1991sw,Baak:2014ora}.
As for the VL quark, which in our scheme eventually decays to light quarks, we take the $T$ mass to be 700~GeV. This value guarantees that the experimental constraints from direct searches \cite{Aad:2015tba} are satisfied. 

\section{Phenomenology of the LHC signal} 
\label{sec:The LHC di-photon excess}

In this Section we determine the values of the Yukawa couplings that yield a large cross section consistent with the current diphoton constraints \cite{ATLAS:2016eeo,CMS:2016crm} and respect upper bounds from the complementary channels obtained with the 8~TeV dataset. We also present a study of the running of the couplings of our model to determine its cutoff scale.

We compute the diphoton cross section in Eq.~\eqref{sppSgg} by evaluating the diphoton decay rate as in Eq.~\eqref{dip}, with $f=E,T$, and the di-gluon decay rate in Eq.~\eqref{dig}. The values of the coupling coefficients $a_f$ used in both the expressions are reported in Appendix~\ref{msmx}. To simplify the phenomenological analysis we set
\be\label{yva}
y_L=y_T\equiv y_v\,,\quad y_N^R=y_N^L.
\ee
and then fix the masses of non-SM particles to the sample values given in Table~\ref{mastab}.
\begin{table}[ht]
\caption{Mass assigments for the three scenarios under consideration.}
\begin{center}
\begin{tabular}{|c|c|c|c|}
\hline
Scenario & I & II & III  \\
\hline
$m_A/{\rm GeV}$  & 330  & 720  & 1000\\
$m_E/{\rm GeV}$  & 165  & 360  & 500\\
$m_{N_1}/{\rm GeV}$  &160 & 345 & 485\\
$m_{N_2}/{\rm GeV}$  &170 & 375 & 515\\
$m_T/{\rm GeV}$  & 700 & 700 & 700\\
\hline
\end{tabular}
\end{center}
\label{mastab}
\end{table}
The three pseudoscalar masses chosen correspond to energies in the diphoton invariant mass distribution showing a one to two $\sigma$ excess over the SM prediction \cite{ATLAS:2016eeo}.
The chosen cross section together with the values of the Lagrangian parameters determined by this choice of masses in the three different scenarios are listed in Table~\ref{partab}. The values of $y_N$ and $y_v$ are determined in each scenario, as explained in Section~\ref{sec:DM}, by matching the observed DM abundance while reproducing the reference diphoton cross section value given in Table~\ref{partab}.
\begin{table}[ht]
\caption{Cross section and parameter values for the three scenarios under consideration.}
\begin{center}
\begin{tabular}{|c|c|c|c|}
\hline
Scenario & I & II & III  \\
\hline
$\sigma_{p p\rightarrow S\rightarrow\gamma\gamma }/{\rm fb}$ & 7.3 & 1.6 & 0.82\\
$m_N/{\rm GeV}$  & 155 & 330 & 470\\
$y_N^R=y_N^L$ & 0.2316 & 0.5907 & 0.6982\\
$y_v$ & 0.400 & 0.421 & 0.448\\
$y_N$ & 0.033 & 0.038 & 0.031\\
$\lambda_{A}$ & 0.020 & 0.030 & 0.038 \\
$\lambda_{AH}$ & 0.200 & 0.390 & 0.505 \\
$\cos\theta$ & 0.015 & 0.02 & 0.015\\
Cutoff scale & $10^{19}$ & $10^{15}$ & $10^{10}$
\\
\hline
\end{tabular}
\end{center}
\label{partab}
\end{table}
We define the mixing angle, expressed in Appendix~\ref{msmx} in terms of the model free parameters, for the VL neutral leptons mass eigenstates by
\be\label{mixN}
N_1=N \cos\theta+N'\sin\theta \,,\quad N_2=N'\cos\theta -N\sin\theta\,.
\ee
The values given in Table~\ref{partab} are such that the lighter neutral mass eigenstate is mostly made of a VL sterile neutrino in each of the three scenarios. 

For each data point we also check that the dijet, $WW$, and $ZZ$ decay cross sections\footnote{In evaluating the $WW$ cross section we make the simplifying assumption that neutral and charged VL leptons be degenerate, which is approximately true in the case at hand, and neglect the contribution of the mostly sterile VL neutrino $N_1$. Furthermore for all the diboson decay rates we assume the electroweak (EW) vector bosons to be massless, given that their mass corrections are of $O\!\left(m_Z^2/m_A^2\right)\sim 1\%$.} satisfy the experimental bounds from the 8~TeV LHC data. Table~\ref{compsearch} shows the values obtained for these quantities and the corresponding upper bounds \cite{CMS:2015neg,Aad:2015agg,Aad:2015kna}. 
\begin{table}[ht]
\caption{Values obtained for the cross sections of complementary channels at 8~TeV, and the corresponding experimental bounds \cite{CMS:2015neg,Aad:2015agg,Aad:2015kna}.}
\begin{center}
\begin{tabular}{|c|cc|cc|cc|}
\hline
  & $\sigma^{\rm th}_{\rm I} ({\rm fb})$ & $\sigma^{\rm exp}_{\rm I} ({\rm fb})$  & $\sigma^{\rm th}_{\rm II} ({\rm fb})$ & $\sigma^{\rm exp}_{\rm II} ({\rm fb})$  & $\sigma^{\rm th}_{\rm III} ({\rm fb})$ & $\sigma^{\rm exp}_{\rm III} ({\rm fb})$ \\
\hline
$jj$  & 3.02  & $-$  & 0.92  & $1.7\times 10^3$  & 1.07  & $860$\\
$WW$ & 0.25 & 690 & 0.033 & 40 & 0.011 & 18 \\
$ZZ$ & 0.029 & 71 & 0.002 & 14 & 0.001 & 10\\
\hline
\end{tabular}
\end{center}
\label{compsearch}
\end{table}

The prediction for the $Z\gamma$ channel in each of the three scenarios is about two orders of magnitude smaller than the corresponding experimental upper constraint on the fiducial cross section \cite{Aad:2014fha}, by definition smaller than the total cross section.\footnote{The fiducial cross section is defined as the total cross section multiplied by the signal acceptance for the given cuts.} The model therefore satisfies the corresponding bound on the $Z\gamma$ total cross section as well. We also calculated the prediction for the invisible decay cross section, respectively equal to 34, 2.6, and 0.66 fb in each of the three scenarios ($m_A=330$, 720, and 1000~GeV), and noticed that the result at $m_A=720$~GeV is about 300 times smaller than the experimental upper bound at 750~GeV quoted in \cite{Franceschini:2015kwy}. This clearly represents a strong indication that the model's predictions for invisible decays in the three scenarios are naturally well within the experimental bounds, though of course only a direct comparison with the relevant experimental bounds can give a definitive answer about the viability of the model's predictions for this channel.

\begin{figure}[t]
  \centering
    \includegraphics[width=.48\linewidth]{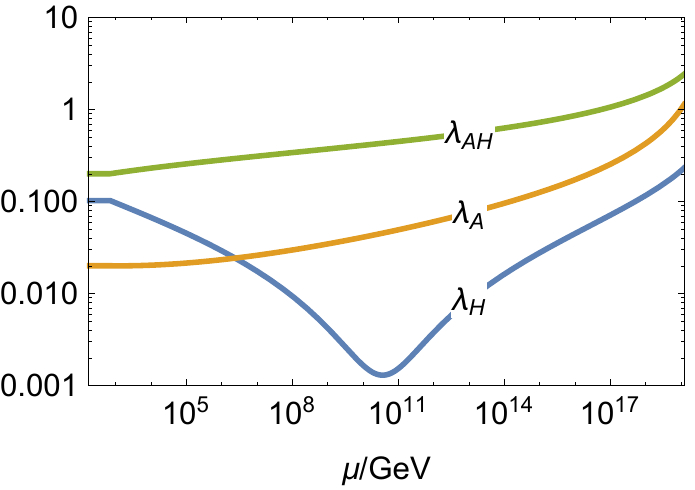}
  \caption{Running of the quartic couplings of the model for scenario I.}
  \label{Run}
\end{figure}

Finally, in Fig.~\ref{Run} we show the running of the quartic couplings for the values of the Lagrangian parameters considered in scenario I, which remains perturbative up to the Planck scale ($10^{19}$~GeV). This value is well above the cutoff scales of $O({\rm TeV})$ obtained for SM-like VL fermions \cite{Franceschini:2015kwy, Bertuzzo:2016fmv}.
In determining the cutoff scale, we required the running couplings to be smaller than $4\pi$ and the scalar potential to be bounded from below. The beta functions of the model are calculated with the help of the {\sc  PyR@TE} package \cite{Lyonnet:2013dna,Lyonnet:2016xiz} and are given in Appendix~\ref{sec:betas}. As customary, we neglect the Yukawa couplings for all SM fermions with the exception of the top quark.

The beta functions of the SM (see for example \cite{Buttazzo:2013uya}) are used to calculate the renormalization group running from the top quark mass $m_t$ up to $330$~GeV. At $m_t$, we take the values of the SM gauge couplings to be $g' = 0.35940$, $g = 0.64754$, $g_3 = 1.1666$, the top Yukawa $y_t = 0.95096$ and the Higgs self-coupling $\lambda_H = 0.12879$ \cite{Buttazzo:2013uya}.

In Table~\ref{partab} we show the values of the cutoff scale for all three scenarios, as well as the corresponding values of the non-SM quartic couplings. The quartic self-coupling of the pseudoscalar and the portal coupling $\lambda_{AH}$ are chosen so as to prevent the Higgs and pseudoscalar self-couplings from running through zero. This is required by the fact that the beta functions of the scalar quartic couplings receive additional negative contributions due to the larger fermion content. Because the running of the gauge and Yukawa couplings is much slower, the model is valid up to a scale where either $\lambda_A$ or $\lambda_{AH}$ become non-perturbative. With these values of parameters, the model can stays perturbative up to the Planck scale in the case of relatively small Yukawa couplings (scenario I) or down to $10^{10}$~GeV for the larger Yukawa couplings of scenario III.

A simple variant of the model can be obtained by assuming flavor universality, which is imposed by assigning negative $\mathbb{Z}_2$ parity to the third quark generation as well. In such a model the mass of the heavy VL quark, $T$, would be bound to be larger than $705$~GeV \cite{CMS:2012vto}: while this value is only slightly above 700~GeV, the chosen $T$ mass, and therefore it does not substantially change the results in Table~\ref{partab}, it is relevant to ask how a larger $m_T$ would affect those results. By using the same assumptions as in Eqs.~\eqref{yva}, the Yukawa coupling $y_v$, and indirectly $y_N^R=y_N^L$ as well, would need to be comparably larger to compensate for the larger $m_T$, given that the diphoton cross section scales as the inverse square of $m_T$. Stability would then require larger pseudoscalar quartic couplings at the $m_A$ scale, which in turn would drive the model to the non-perturbative regime at a scale lower than the one given in Table~\ref{partab}.

\section{An authentic WIMP candidate} 
\label{sec:DM}

We investigate now the compatibility of the measured DM abundance with the relic abundance of the VL neutrino $N_1$.

DM provides 26\% of the energy density of the present Universe. As mentioned in Section~\ref{intro}, DM is usually modelled after a WIMP because particles having masses and annihilation cross sections set by the EW scale provide the measured value of DM abundance in a natural way~\cite{Jungman:1995df, Bertone:2004pz}.

Our model indeed presents a natural candidate for DM, the VL neutrino $N_1$ that, owing to its SM weak interactions emerges as an authentic WIMP.\footnote{The heavier VL neutrino $N_2$ eventually decays to $N_1$ by emitting SM light-fermion pairs via virtual $Z$ bosons.} We calculate the relic abundance of $N_1$, arising from the standard freeze-out scenario,
\begin{equation}
	\Omega_{N_1}h^2 \simeq 0.1 \sqrt{\frac{60}{g_\star(T_{\rm fo})}} \frac{\langle \sigma v \rangle_{\rm fo}}{\langle \sigma_{N_1\bar{N}_1 \to SM} v \rangle(T_{\rm fo})},
\end{equation}
where $g_\star(T_{\rm fo})$ is the effective number of relativistic degrees of freedom at the freeze-out temperature $T_{\rm fo}$, $\langle \sigma v \rangle_{\rm fo} = 3 \times 10^{-27}$ cm$^3$s$^{-1}$ is the standard freeze-out cross-section and $\langle \sigma_{N\bar{N} \to SM} v \rangle(T_{\rm fo})$ is the velocity averaged annihilation cross-section of $N_1$ to SM particles at the freeze-out temperature $T_{\rm fo}$. For the latter, we assume the dominance of VL neutrino annihilation mediated by an $s$-channel pseudoscalar $A$ into gluons, enhanced by assuming a VL neutrino mass close to the threshold. Other annihilation channels,  like the $s$-channel annihilation through $Z$ into a pair of SM particles, are highly suppressed, and therefore negligible, given that the VL neutrino is at 99\% made of a sterile neutrino gauge eigenstate.

\begin{figure}[ht]
  \centering
    \includegraphics[width=0.75\linewidth]{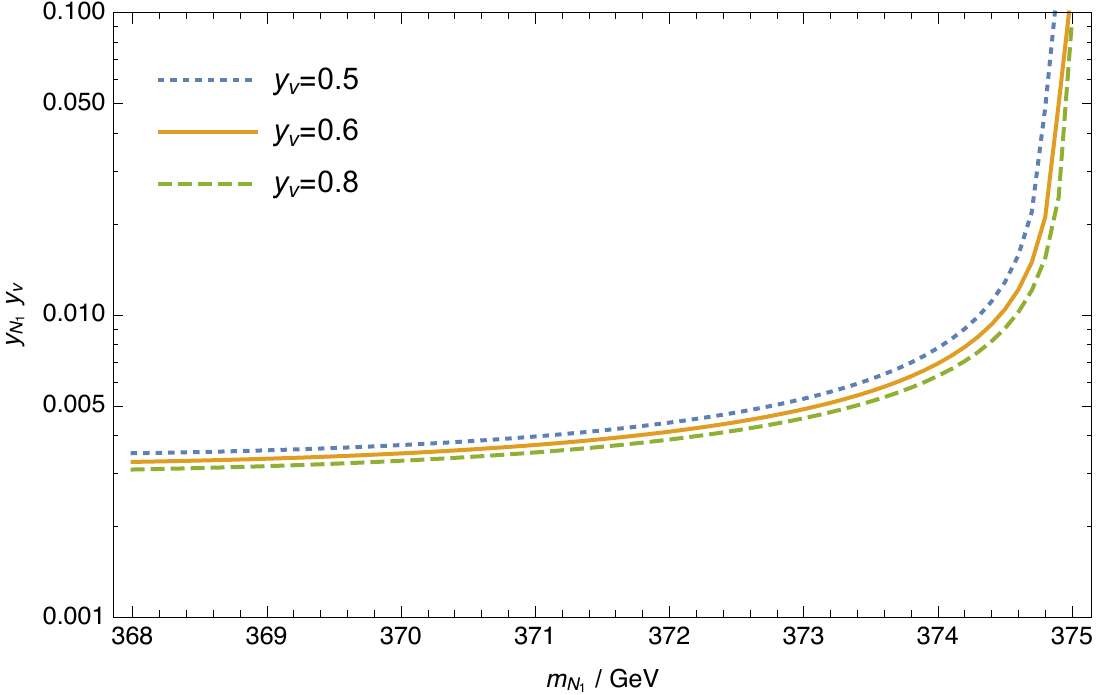}
  \caption{Value of $y_{N_1}\times y_v$ producing the observed DM relic abundance, as measured by the Planck collaboration~\cite{Ade:2015xua}, for a range of lighter VL neutrino masses near the resonance and for $y_v=0.36,0.4,0.44$.}\label{fig:DM}
\end{figure}

For a lightest VL neutrino with mass $m_{N_1}$, whose value for each scenario is given in Table~\ref{DMtab}, we find the value of the physical coupling to $A$, $y_{N_1}$, reproducing the DM relic abundance as measured by the Planck collaboration, $\Omega_{\rm DM} h^2 = 0.1188 \pm 0.0010$~\cite{Ade:2015xua}.
\begin{table}[ht]
\caption{DM candidate mass and coupling to pseudoscalar mediator.}
\begin{center}
\begin{tabular}{|c|c|c|c|}
\hline
Scenario & I & II & III  \\
\hline
$m_{N_1}/{\rm GeV}$  &160 & 345 & 485\\
$y_{N_1}$ & 0.019 & 0.020 & 0.017\\
\hline
\end{tabular}
\end{center}
\label{DMtab}
\end{table}
By using for each scenario the value of $y_{N_1}$ listed in Table~\ref{DMtab} in combination with Eq.~\eqref{yA}, we can express $y_N$ as a function of $y_v$, with the remaining parameters given in Tables~\ref{mastab} and \ref{partab}, and determine $y_v$ subsequently by matching the observed diphoton cross section. 
In Fig.~\ref{fig:DM} we plot the value of $y_{N_1}$ producing the observed DM relic abundance for a range of lighter VL neutrino masses near the resonance and for $y_v=0.36,0.4,0.44$.

The relic abundance of $N_1$ is strongly constrained by the direct detection experiments, which put an upper bound on the DM elastic scattering off a nucleon. In the case at hand, the only parton that interacts with the pseudoscalar mediator is the gluon, and the only effective operator relevant for such process can be written as 
\be\label{GGNN}
{\cal O}_{g{\operatorname{-}}N_1}=\frac{1}{\Lambda_{F}^3}\,i\bar{N}_1\gamma_5 N_1\cdot
\frac{\alpha_S}{8\pi} G^{a\mu\nu} \tilde{G}^a_{\mu\nu}\,,
\ee
with $\Lambda_F$ being the effective scale of the interaction. The resulting cross section is both momentum suppressed and spin-dependent, in which case the current experimental upper bound is still rather large \cite{Akerib:2016lao} and therefore not yet sensitive enough to constrain the theory \cite{Chu:2012qy}.

\section{Conclusions}
\label{sec:Conclusions}

While the preliminary evidence for the 750 GeV excess at the LHC has not be confirmed by the latest results that both ATLAS and CMS experiments presented at ICHEP 2016, the small background of the diphoton channel and its sensitivity to new physics via its leading order loop coupling make it a privileged experimental test for new physics. In this spirit, we took the initial evidence in favor of a 750~GeV resonance as a template to study possible weakly-interacting new physics contributions in conjunction with the appearance of high-mass diphoton resonances. The main result of this paper is to demonstrate that a diphoton excess can be enhanced by threshold effects due to new light particles, that evade direct search current constraints. By studying three different scenarios we show that a large excess in the diphoton spectrum can be explained within weakly coupled theories that retain perturbativity up to energy scales as large as the Planck scale. 

Our results rely on a prototypical model containing a vector-like EW doublet and singlet leptons and a vector-like top quark, as summarized in Table~\ref{4thgen}. In addition to satisfying the LHC upper bounds on possible large diphoton excesses in the invariant mass spectrum and the bounds of complementary channels, our framework proposes a natural dark matter candidate. This is the lighter vector-like neutrino, whose relic abundance accounts for the entire measured dark matter abundance in the same part of the parameter space selected by the speculated LHC signals. The direct and indirect DM detection constraints are also satisfied in the same parameter ranges.

\section*{Note added}
While the present paper was being finalized and several of the authors were attending the EW session of the 2016 Moriond conference, a paper
using a threshold enhancement of the diphton cross section similar to that studied in our paper appeared on the arXiv \cite{Bharucha:2016jyr}. 
We remark that our model differs in particle content and interactions by the one of \cite{Bharucha:2016jyr}; we furthermore show that, while  satisfying the observed 8~TeV constraints on the diphoton and other diboson cross sections, our model can remain perturbative up to the Planck scale in the best case.

\section*{Acknowledgement}

This work was supported by the Estonian Research Council grants PUTJD110, PUT808 and PUT799, the grant IUT23-6 of the Estonian Ministry of Education and Research, and by the EU through the ERDF CoE program grant TK133. AH thanks the Horizon 2020 programme as this project has received funding from the EU Horizon 2020 programme under the Marie Sklodowska-Curie grant agreement No 661103.

\appendix
\section{Vector-like fermion masses and couplings}
\label{msmx}
The masses of the VL neutral leptons $N_{1,2}$, defined by Eq.~\eqref{mixN}, are
\be\label{mx12}
m_{N_{1,2}}=\frac{1}{2} \sqrt{l^2+L^2+m^2+M^2\mp2 \sqrt{\left(l^2+m^2\right) \left(L^2+M^2\right)}}\,,
\ee
with  mixing angle determined by
\be
\tan2\theta=\frac{l M-L m}{l L+m M}\,,
\ee
where
\be\label{MmLl}
M=m_L+m_E\,,\ m=m_E-m_L\,,\ L=\frac{y^R_N+y^L_N}{\sqrt{2}}v_w\,,\ l=\frac{y^R_N+y^L_N}{\sqrt{2}}v_w \,.
\ee
Finally, the coupling coefficients of the mass eigenstates $E$ and $T$ appearing in Eqs.~(\ref{dip},\ref{dig}) are simply
\be
a_E=-\frac{y_L v_w}{\sqrt{2} m_L}\,,\quad a_T=-\frac{y_T v_w}{\sqrt{2} m_T}\,,
\ee
while the Yukawa coupling of the physical VL neutrinos $N_{1,2}$ to $A$ are
\be\label{yA}
y_{N_{1,2}}=\frac{ 2 y_{N} \left(m^2 M^2\mp m M\sqrt{\left(l^2+m^2\right) \left(L^2+M^2\right)}\right)+l^2 M^2 \left(y_{N}+y_L\right)+L^2 m^2
   \left(y_{N}-y_L\right)}{4 \left(l^2+m^2\right) \left(L^2+M^2\right) }\frac{{\cal N}_{1,2}}{\sqrt{2}}
   \ee
with
\be
{\cal N}_{1,2}=\sqrt{1+\left|\frac{l L-m M\mp \sqrt{\left(l^2+m^2\right) \left(L^2+M^2\right)}}{L m+l M}\right|^2} \sqrt{1+\left|\frac{l L+m M\pm
   \sqrt{\left(l^2+m^2\right) \left(L^2+M^2\right)}}{L m-l M}\right|^2}\,.
\ee

\section{Beta functions}
\label{sec:betas}
The one-loop beta functions for our model, calculated with the {\sc PyR@TE} package \cite{Lyonnet:2013dna,Lyonnet:2016xiz}, are given by:
\begin{align}
  16 \pi^2 \beta_{g_1} &= \frac{167}{18} g^{\prime 3},
  \\
  16 \pi^2 \beta_{g} &= -\frac{5}{2} g^3,
  \\
  16 \pi^2 \beta_{g_3} &= -\frac{19}{3} g_3^3,
  \\
  16 \pi^2 \beta_{y_t} &= y_t \left[ \frac{9}{2} y_t^2 + 12 (|y_L^L|^2 + 3 |y_Q^L|^2 
  + 3 |y_Q^R|^2) - \frac{17}{12} g^{\prime 2} 
  - \frac{9}{4} g^2 - 8 g_3^2 \right],
  \\
16 \pi^2 \beta_{\lambda_H} &= \frac{3}{8} (g^{\prime 4} + 2 g^{\prime 2} g^2 + 3 g^4) - 6 y_t^4 + 24 \lambda_H^2 + 2 \lambda_{AH}^2 - 2 |y_N^L|^4 - 2 |y_N^R|^4 
 \notag
 \\
 & + \lambda_H (-3 g^{\prime 2} - 9 g^2 + 12 y_t^2 + 4 |y_N^L|^2 + 4 |y_N^R|^2),
  \\
  16 \pi^2 \beta_{\lambda_A} &= 2 \left[ 36 \lambda_{A}^2 + \lambda_{AH}^2 - 2 |y_L|^4 - |y_N|^4 - 
   3 |y_T|^4 + 4 \lambda_{A} (2 |y_L|^2 + |y_N|^2 + 3 |y_T|^2) \right],
 \\
   16 \pi^2 \beta_{\lambda_{AH}} &= 8 \lambda_{AH}^2 + \lambda_{AH} \left[-\frac{3}{2}( 3 g^2 + g^{\prime 2})   + 6 y_t^2 +  24 \lambda_{A} + 12 \lambda_{H} + 8 |y_L|^2 + 4 |y_N|^2 
   \right. \notag
   \\
   & \left.
   + 2 |y_N^L|^2 + 2 |y_N^R|^2 + 12 |y_T|^2 \right] - 
 4 (|y_L|^2 |y_N^L|^2 + |y_N|^2 |y_N^L|^2 + 
    |y_L|^2 |y_N^R|^2 + |y_N|^2 |y_N^R|^2 
    \notag
    \\ 
   & + y_N^L y_N^R y_L^{*} y_N^{*} + 
    y_L y_N y_N^{L *} y_N^{R *}), 
    \end{align}
    \begin{align}
  16 \pi^2 \beta_{y_N^L} &= \frac{1}{4} y_N^L \left[ -3 (3 g^2 + g^{\prime 2} - 4 y_t^2) + 
    2 (|y_L|^2 + |y_N|^2 + 5 |y_N^L|^2 + 2 |y_N^R|^2) \right] + 
 2 y_L y_N y_N^{R *},
 \\
 16 \pi^2 \beta_{y_N^R} &= \frac{1}{4} y_N^R \left[-3 (3 g^2 + g^{\prime 2} - 4 y_t^2) + 
    2 (|y_L|^2 + |y_N|^2 + 2 |y_N^L|^2 + 5 |y_N^R|^2) \right] + 
 2 y_L y_N y_N^{L *},
 \\
 16 \pi^2 \beta_{y_L} &= \frac{1}{2} y_L (-9 g^2 - 3 g^{\prime 2} + 14 |y_L|^2 + 4 |y_N|^2 + |y_N^L|^2 + 
    |y_N^R|^2 + 12 |y_T|^2) + 2 y_{N}^{L} y_{N}^{R} y_{N}^{*},
 \\
 16 \pi^2 \beta_{y_N} &= y_{N} (4 |y_L|^2 + 5 |y_N|^2 + |y_N^L|^2 + |y_N^R|^2 + 
    6 |y_T|^2) + 4 y_{N}^{L} y_{N}^{R} y_{L}^{*},
\\
 16 \pi^2 \beta_{y_T} &= y_{T} (-8 g_{3}^2 - \frac{8}{3} g^{\prime 2} + 4 |y_L|^2 + 2 |y_N|^2 + 9 |y_T|^2).
\end{align}

\bibliographystyle{hunsrt}
\bibliography{bib.bib}

\end{document}